\documentclass[aps,pre, floatfix,10pt]{revtex4-2}
\usepackage{graphicx}
\usepackage{dcolumn}
\usepackage{bm}
\usepackage{latexsym}
\usepackage{amsmath}
\usepackage{amssymb}
\usepackage{color}
\usepackage[normalem]{ulem}
\usepackage{multirow}
\begin{document}
\title{{\bf Flagellar length control in multiflagellated eukaryotes: a case study with {\it Giardia}}}
\author{Swayamshree Patra} 
\affiliation{Department of Physics, Indian
  Institute of Technology, Kanpur 208016, India} 
\author{Debashish Chowdhury{\footnote{E-mail: debch@iitk.ac.in}}}
\affiliation{Department of Physics, Indian Institute of Technology, 
  Kanpur 208016, India}

\begin{abstract}
Every organism has a size that is convenient for its function. Not only multicellular organisms but also uni-cellular organisms and even subcellular structures have convenient sizes.  Flagella of eukaryotic cells are long dynamic cell protrusions. Because of their simple linear geometry, these cell appendages have been popular system for experimental investigation of the mechanisms of size control of organelles of eukaryotic cells. In the past most of the attention have been focussed on mono-flagellates and bi-flagellates. By extending our earlier model of bi-flagellates, here we develop a theoretical model for flagellar length control in {\it Giardia} which is an octo-flagellate. It has four pairs of flagella of four different lengths. Analyzing our model we predict the different sizes of the four pairs of flagella . This analysis not only provide insight into the physical origins of the different lengths but the predicted lengths are also consistent with the experimental data.

\end{abstract}

\maketitle

\section{Introduction}

Size matters. ``Being the right size'' \cite{haldane25} ensures proper biological function of a living organism. In fact, not only the size of a multi-cellular organism \cite{bonnerbook}, but also uni-cellular organisms and even subcellular compartments (organelles) are known to have sizes that are believed to have been optimized by evolution for their respective functions \cite{marshall16,marshall02,marshall15,goehring12}.  Among the subcellular structures of eukaryotic cells there are some that are effectively linear \cite{marshall02,folz19,gov06,orly14,renault17}. Because of their one-dimensional geometry, the problem gets simplified to that of length control of a linear appendage of the cell. One of these organelles, called cilium, which is also referred to as flagellum (not to be confused with bacterial flagellum) \cite{ginger08}, is the most popular system for the studies of subcellular length control. We use the terms cilia and flagella interchangeably. In this paper we focus exclusively on the mechanisms of length control of flagella of eukaryotic cells.

The following are some common features and questions regarding the  growth, development and maintenance which are shared by the flagella of eukaryotic cells.  First, note that  wild type cells assemble their  flagella  from scratch \cite{cavaliersmith74,cross15,quarmby05}. The flagella continue elongating with a rate that decreases with increasing length and this process of ciliogenesis stops after the elongating flagella attain the desired length. Such a control of the rate of growth of a flagellum on its instantaneous length indicates the existence of a length-sensing mechanism of the cell. So, the first question is: what is the identity of the ruler (or timer) that the cell uses to measure length and what is the mechanism of measuring the instantaneous length of the flagellum using this ruler (or timer) \cite{ludington15}? 

Second, flagella elongate by adding its constituent structural proteins at its distal tip. But, these proteins (referred to as ``precursor'', from now onwards in this paper) are not synthesized in the flagella. Instead, these proteins are synthesized in the cell body and then transported along its tubular chamber to the tip where these are incorporated  into the flagellar structure \cite{kozminski93,kozminski12,rosenbaum02}. So, the second question is: how does the cell exploit the information on the flagellar length to regulate the intraflagellar transport (IFT) of the proteins that results in a length-dependent rate of the assembly? Is regeneration of flagella to their original length possible if the cell either resorbs or sheds off the original pair of its flagella?

Third, even after a flagellum attains its steady-state length, assembly and disassembly of structural proteins at the tip continues. Because of the turnover of tubulins at the flagellar tip, the flagellum would shorten by removal of  precursors unless the shortening is compensated by equal amount of elongation by fresh assembly at the tip \cite{marshall01,lechtreck17,ludington15,ishikawa17}. Thus, compensation of elongation and shortening caused by these two competing processes maintains the average length constant at this `balance point' \cite{marshall01,mirvis18}. In order that such a balance point exists at least one of these two rates must be length-dependent. So, the third question is: what is the explicit form of length-dependence of the rates of assembly and/or disassembly? 

Fourth, in case of multiflagellates or monoflagellates which become transiently multiflagellated \cite{patra21},  the multiple flagella coordinate their growing and shortening dynamics.  So, the fourth question is: how do the two flagella communicate with each other through a common shared pool of resources like structural proteins and transport vehicles \cite{fai19,patra20}? If one of the flagella is amputated selectively, how does the other flagellum respond to this damage during the regeneration of the amputated flagellum \cite{patra20}? 

Significant progress have been made in the last few years in the search of answers to the questions posed above. Answer to these questions are known in the context of the biflagellate {\it Chlamydomonas reinhadtii} which posses a pair of flagella of same length \cite{wemmer07}. However, the problem becomes  more complex in the case of {\it Giardia} cell which is an octoflagellate eukaryote and possesses four pairs of flagella of four different lengths \cite{mcinally19}. To our knowledge, this eukaryote has received very little attention so far from the perspective of length control of its eight flagella. In this paper we develop a theoretical model for the elongation/shortening kinetics of the flagella of {\it Giardia} and report new results. We show that our results are consistent with the experimental data reported earlier in the literature. This model can guide the analysis of future experiments on the kinetics of the flagella of {\it Giardia} in further detail.

\section{Features of flagella and IFT in Giardia cell}

\begin{figure*}
\includegraphics[width=0.70\textwidth]{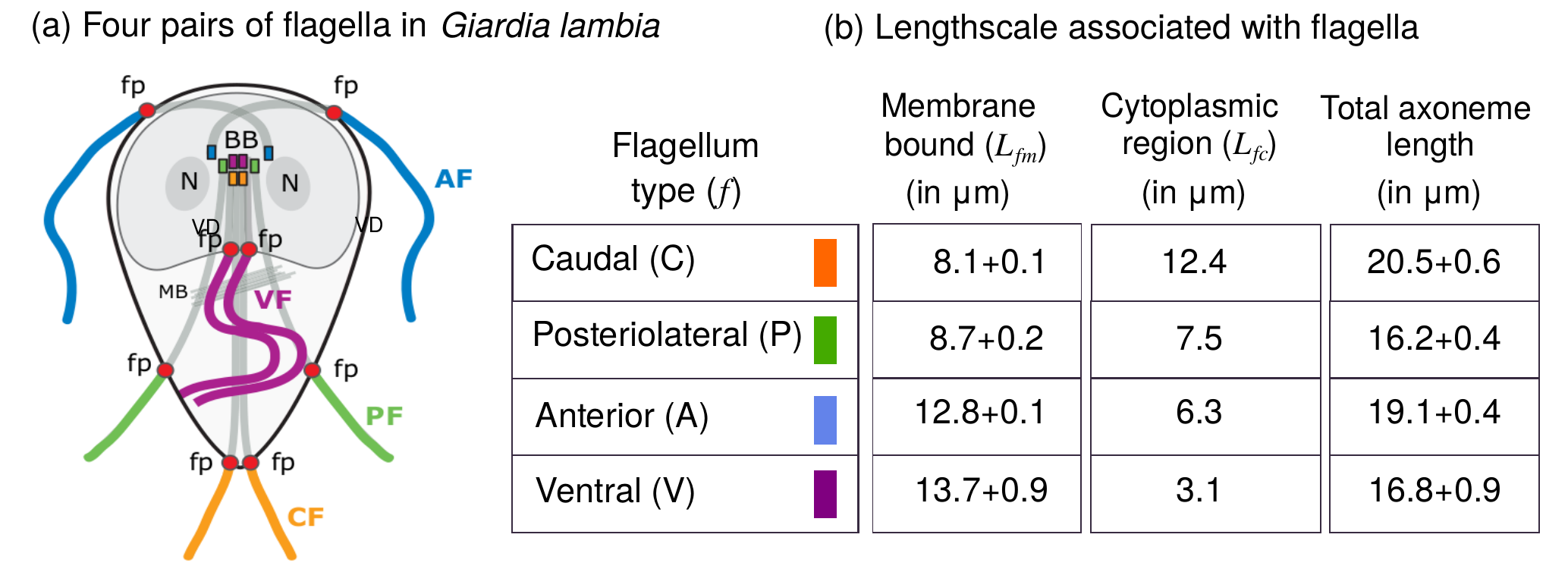}
\caption{{ {\bf Giardia cell with four pair of flagella: } (a) ({Reproduced from ref.\cite{mcinally19}, which was published by McInally et al. \cite{mcinally19} under {\it Creative Commons Attribution License}}). Four pair of flagella of a {\it Giardia } cell are named as caudal(C), ventral(V), anterolateral(A) and posteriolateral (P). Axoneme of each flagellum has a membrane bound region and a region exposed in the cytoplasm. The initiating point of the axoneme is the basal body (BB) and the point through which axoneme protrude the cell membrane is known as the flagellar pore (fp). The other microtubule based structures are the median body (MB) and the ventral disc (VD). The pair of nuclei are denoted by N. . (b) In the table, the membrane bound length of the axoneme ($L_{fm}$) and the length of the axoneme exposed in the cytoplasm  ($L_{fc}$)  for flagellum $f$ are tabulated. Data originally reported in ref.\cite{mcinally19}.}}
\label{fig_giardia}
\end{figure*}

{\bf Flagellar length:} An axoneme is at the core of the structural scaffold of a flagellum. The axoneme of a flagellum is assembled on a basal body and projects out towards, and beyond, the cell surface. The major structural components of the axoneme are microtubule (MT) doublets. Each MT is a tubular stiff filament. In a 9+2 axoneme nine  microtubule (MT) doublets are arranged cylindrically with the two coaxial central pair. The four pairs of flagella of a {\it Giardia} cell are referred to as caudal (C), ventral (V), anterior (A) and  posteriolateral (P) as depicted in Fig.{\ref{fig_giardia}(a)}. Unlike the other flagellates, the axoneme of  {\it Giardia} is very long and has both a membrane bound region and a region exposed in the cytoplasm. For a given flagellum $f$ in a {\it Giardia} cell, there are three lengths of interest: (i) the length of the axoneme exposed in the cytoplasm $L_{fc}$, (ii) membrane bound axoneme length $L_{fm}$ and (iii) total length of the axoneme $L_{fm}+L_{fc}$. For all the four pair of flagella, the magnitudes of these lengths are tabulated in Fig.\ref{fig_giardia}(b).

\begin{figure*}
\includegraphics[width=0.60\textwidth]{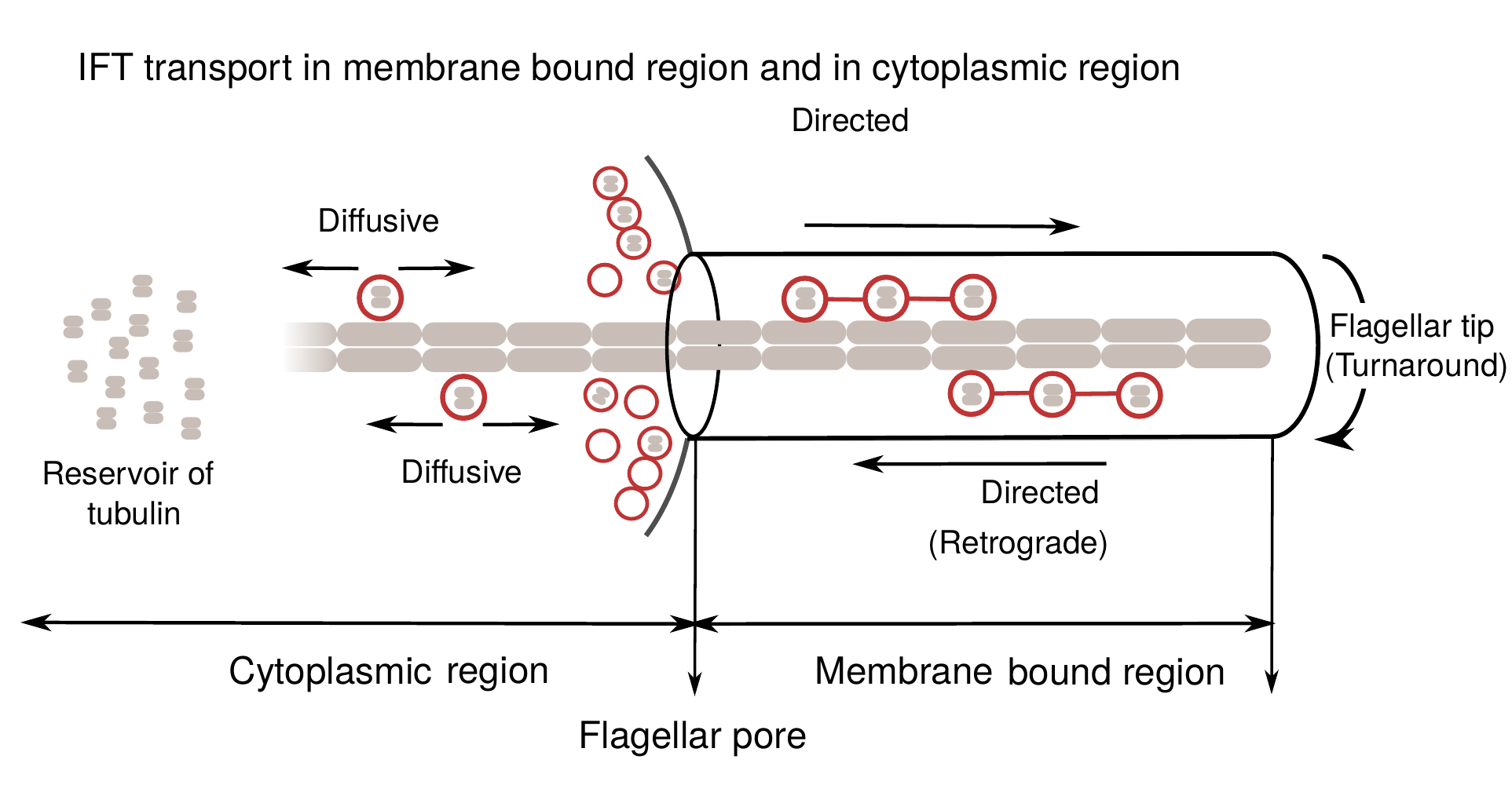}
\caption{{ {\bf Intraflagellar transport in {\it Giardia cell}: } Independent IFT particles are observed to perform diffusive movement in the region of axoneme lying exposed in the cytoplasmic region. In the membrane bound region,
the IFT particles assemble to form long IFT trains and shuttle inside the flagellum in a directed fashion. One filament of the microtubule doublet  is utilised for the journey from the flagellar pore to the tip whereas the other one is used for the trip back from the tip to the pore. Tubulins from the microtubule structures inside the cell body like median body and ventral disc are released by the depolymerase motors kinesin-13 and carried to the local pools at the flagellar pore by the diffusing IFT particles.}}
\label{fig_ift}
\end{figure*}

{\bf Intraflagellar transport:} The MTs serve as tracks for two distinct `superfamilies' of molecular  motors, called kinesin and dynein. Each MT doublet supports two-way transport: anterograde (base-to-tip) transport of cargo is driven by kinesin motors on one of the two MTs while the retrograde (tip-to-base) transport is driven by dynein motors on the other MT \cite{stepanek16}. Directed movement of the motors belonging to both the families is powered by ATP hydrolysis. This phenomenon of intraflagellar transport (IFT) \cite{kozminski93,kozminski12,rosenbaum02} is required not only for ciliogenesis and regeneration of flagella but also for maintaining a constant average length of the flagellum in ther steady-state. The cargo trains that are hauled by kinesins and dyneins are called IFT trains and actually linear assemblies of proteins called IFT particles \cite{lechtreck17,prevo17,wren13,craft15}. The tubulin subunits of MTs are the main molecular cargo that are loaded onto the IFT trains for transport and delivery at the target locations. The IFT particles perform diffusive motion on the axoneme in the cytoplasmic region. They are assembled as trains at the flagellar pore and perform directed motion in the membrane bound part. The tubulin released from the depolymerization of microtubules of the median body and ventral disc are utilised for assembling the membrane bound flagellum. From these locations, the tubulins are transported to the flagellar pore by the diffusive IFT particles and from the pore, they are transported by the directed IFT trains to the flagellar tip. A local pool of tubulin and IFT particles is maintained near the flagellar pore. See Fig.\ref{fig_ift} for  a schematic description of the IFT in {\it Giardia } cell.

\begin{figure*}
\includegraphics[width=0.60\textwidth]{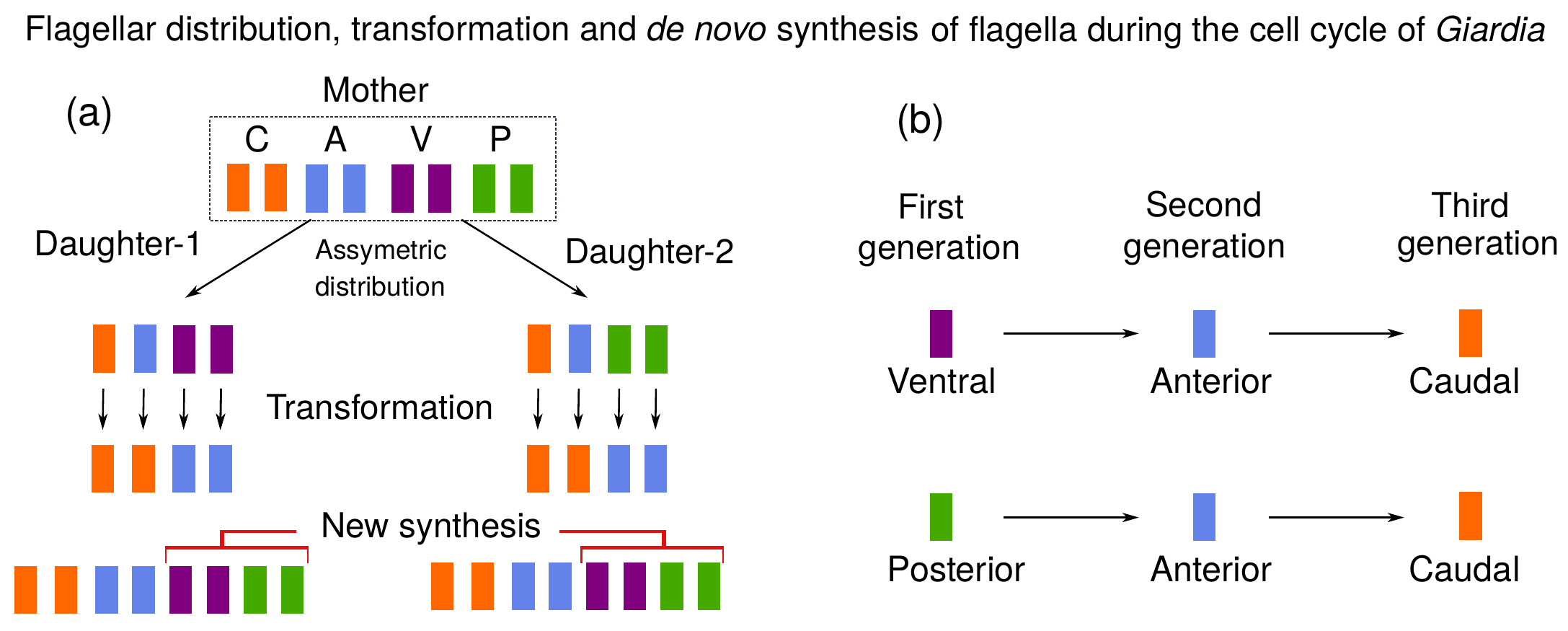}
\caption{ {\bf Flagellar distribution during cell division of {\it Giardia cell}}: (a) Two daughters receive different set of flagella from mother cell. Each of the daughter cell receive one caudal and an anterior flagellum. One of them receives a pair of ventral flagella and the other receives a pair of posteriolateral flagella. In the daughter cell which receives a caudal, an anterior and a pair of ventral flagella, the anterior flagellum gets transformed into caudal flagellum, the pair of ventral flagella transform into a pair of anterior flagella and the cell assembles a new pair  of both ventral and posteriolateral flagella. Similarly, in the other daughter cell which receives a caudal, an anterior and a pair of posteriolateral flagella, the anterior flagellum gets transformed into caudal flagellum, the pair of posteriolateral flagella transforms into a pair of anterior flagella and the cell assembles a new pair of both ventral and posteriolateral flagella. (b) The transformation of the ventral and posteriolateral flagella over multiple cell cycle.   }
\label{fig_division}
\end{figure*}

{\bf Cell cycle and flagella distribution:} Another interesting feature of {\it Giardia} is the transformation of flagella during cell division. As shown schematically in Fig.{\ref{fig_division}}(a), two daughters initially receive different set of flagella i.e, the distribution is asymmetric. Each of the daughter cell receive one caudal and an anterior flagellum. One of them receives a pair of ventral flagella and the other receives a pair of posteriolateral flagella making the distribution. In the daughter cell which receives a caudal, an anterior and a pair of ventral flagella, the anterior flagellum gets transformed into caudal flagellum, the pair of ventral flagella transforms into a pair of anterior flagella and the cell assembles a new pair  of both ventral and posteriolateral flagella. Similarly, in the other daughter cell which receives a caudal, an anterior and a pair of posteriolateral flagella, the anterior flagellum gets transformed into caudal flagellum, the pair of posteriolateral flagella transforms into a pair of anterior flagella and the cell assembles a new pair of both ventral and posteriolateral flagella \cite{nohynkova06}. In Fig.\ref{fig_division}(b) it can be seen how over multiple cell cycles the flagellum transforms from one type to another.

{\bf Length control mechanism:} The question is, whether the cell measures the length of the membrane bound axoneme only or the length of the whole axoneme. In a recent paper, the authors claimed that the pair of caudal flagella has the shortest membrane bound length \cite{mcinally19}. They observed that the amount of depolymerase at the tip of caudal flagella is the highest as compared to that in the other flagella. As a result, the membrane bound length of the caudal flagellum is the shortest as compared to the other flagella. Hence, length of the membrane bound flagellum is inversely proportional to the amount of depolymerase motors at the tip. In order to check whether the uneven distribution of kinesin-13 could drive the emergence of different steady state lengths, a mathematical model was developed to complement the experimental observation \cite{mcinally19}. 



\section{Generic model for controlling the membrane bound flagellar length}

In this section we summarize the general modeling strategy that we have developed recently for flagellar length control and applied to monoflagellates \cite{patra21} and biflagellates \cite{patra20}. 
In this approach, a flagellum is represented as a pair of antiparallel lattices, where each lattice represents a MT of the doublet. For both monoflagellates and biflagellates that were modelled in refs.\cite{patra21,patra20}, $L_{fc} \simeq 0$ so that $L_{f} \simeq L_{fm}$, i.e., the length of the membrane-bound part of each flagellum exposed outside the cell is almost equal to that of the entire flagellum.
 
Three different types of proteins perform their respective distinct key functions in IFT. (i) `Precursor' is a generic term used for proteins that are essential components of the structure of the  flagellum. Since the core of that structure is formed by the axoneme and since tubulins are the key constituents of axoneme the precursors can be identified with tubulins. (ii) The precursors are dispatched to the flagellar tips by loading onto IFT particles which are also proteins and can assemble into IFT trains. As explained below, not all the trains necessarily carry cargo during the journey from the flagellar base. (iii) IFT trains, irrespective of whether loaded with cargo or not, are pulled by molecular motors that walk along their respective MT tracks along the narrow space between the axoneme and the flagellar membrane. The number of motors per IFT train is not known. Therefore, instead of describing the motion of the motors explicitly, these models describe the directed movement of the entire composite macromolecular complex formed by an individual IFT particle and the motors pulling it. Each such composite object is represented by a self driven particle.

In the stochastic formulation of the kinetics of the model, a particle can probabilistically hop  to the next site in front, with a certain probability per unit time (hopping rate), obeying an exclusion principle: no site can be occupied simultaneously by more than one particle. However, because of the validity of a time-scale separation  \cite{patra20}, the traffic of the IFT particles can justifiably be assumed to be in a steady-state. The symbols $\rho$, $v$ and $J$ denote the number density on the lattice, average velocity and flux of the IFT particles inside the flagellum in the steady-state.

Each IFT particle can be either empty or loaded with a single precursor (lattice unit) which, upon arrival at the flagellar tip,  can elongate the model flagellum (pair of lattices) at the tip by a single unit.  
Whether or not an IFT particle is to be loaded with a precursor before it begins its journey from the flagellar base depends on the length of the flagellum at that instant of time \cite{wren13,craft15}.
In this model the cell is assumed to sense the flagellar length with a {\it time of flight} mechanism. Suppose, a timer molecule, which is an integral part of an IFT train, enters the flagellum in a particular chemical (or conformational) state. The probability of finding the timer in the same state after time $t$ is proportional to $e^{-k t}$ where $k$ is rate of irreversible flipping of the state of the two-state timer.  So, if the mean flagellar length is $L_{f} (t)$ at time $t$, the probability of the timer returning to the base (after the roundtrip inside flagellum) without flipping its chemical state is proportional to $e^{-k t_{tof}}$ where $t_{tof}$ is the time of flight and is given by $t_{tof}=2 L_{f}(t)/v$. Hence, if the timer returns without flipping its state, the IFT train loaded with a precursor is dispatched, otherwise empty IFT train enters the flagellum \cite{wren13}. Whether the IFT train is loaded or not with a precursor also depends on the amount of precursor in the pool. If the current average  population of pool is $n (t)$ and its maximum capacity  is $n_\text{max}$, then the probability of loading of precursors onto an IFT train is proportional to $n(t)/n_\text{max}$. Therefore, the flux of loaded trains reaching the flagellar tip is
\begin{equation}
J_{full}=\frac{n(t)} {n_\text{max}} ~ e^{-2k L_{f}(t) /v} J~.
\end{equation}

Successful incorporation of a precursor, with probability $\Omega_e$, into the flagellar tip would elongate the flagellum. So the overall rate of assembly is given by
\begin{equation}
{\text{Rate of elongation}} = \frac{n(t)}{n_\text{max}} J \Omega_e ~ e^{-2k L_{f}(t) /v}.
\label{eff-ass-rate}
\end{equation} 
Because of the ongoing turnover, if both the sites at the tip are not occupied by any IFT trains, the dimer (i.e the precursor) dissociates with rate $\Gamma_r$. Therefore, the overall dis-assembly rate is given by
\begin{equation}
{\text{Rate of shortening}} = (1-\rho)^2 \Gamma_r~.
\label{eff-dis-rate}
\end{equation} 
Some plausible candidates, that can serve as timer, have been considered although conclusive identification remains to be established. It could be either a small molecule bound to the IFT particle or a small segment of the IFT particle itself.

The pool synthesizes and degrades  precursors in a population-dependent fashion so that precursor population never  exceeds $n_\text{max}$.  Thus, the rate of synthesis $\omega^+\left({n(t)} \right)$ and that of degradation $\omega^-\left({n(t)} \right)$ are dependent on the pool population $n(t)$. The coupled set of differential equations which govern the evolution of {\it mean} flagellar length $L_{f}(t)$ and pool population $n(t)$ are 
\begin{eqnarray}
\frac{dL_{f}(t)}{dt} &=& \frac{n(t)}{n_\text{max}}J\Omega_e \exp \left(  -\frac{2k L_{f}(t)}{v} \right) -(1-\rho)^2 \Gamma_r \nonumber \\
\frac{dn(t)}{dt}&=&\omega^+ \left({n(t)} \right) - \omega^- (n(t)) -\frac{dL_{f}(t)}{dt}~.
\label{flagella-rate-eq}
\end{eqnarray}

\section{Model of flagellar length control specifically in Giardia}

As explained above, a unique feature of the flagella of {\it Giardia} is that, for a given flagellum, there are two distinct lengths of primary interest: $L_{fc}$ and $L_{fm}$. One interesting pattern observed is that the axoneme length exposed in the cytoplasmic region is longest for the pair of caudal flagella which have the shortest membrane bound axoneme length whereas the axoneme length exposed in  the cytoplasmic region is shortest for the pair of ventral flagella which have the longest membrane bound axoneme length (see Fig.\ref{fig_giardia}(b)). 

Kinesin-13 is localised uniformly on the interphase microtubule structures such as ventral disc and median body. It is possible that kinesin-13 is liberating tubulin by its depolymerising activity from the microtubules forming the ventral discs and median body and hence responsible for supplying free tubulins for the local precursor pool near the flagellar pore. IFT particles are known to diffuse on the cytoplasmic axoneme {\cite{mcinally19} } and may be responsible for carrying the released tubulins from the microtubule based structures to the flagellar pores of different flagella. Near the flagellar pores, a local pool of tubulin must be maintained for assembling and maintaining the membrane bound axoneme. Let $\omega^+_{f}$ and $\omega^-_{f}$ denote, respectively, the rates of accumulation and depletion of tubulins in the local precursor pools near the flagellar pore of flagellum labelled by $f$. 

If the length of the cytoplasmic axoneme is $L_{fc}$, the typical time taken by a diffusing IFT particle to reach the flagellar pore is 
\begin{equation}
\tau=\frac{L_{fc}^2}{2D}
\end{equation}
So the rate of addition of precursors to the local pool of flagellelum $f$ is 
\begin{equation}
\omega^+_{f}=\alpha^+ \frac{D}{L_{fc}^2}
\end{equation} 
where $\alpha^+$ is a constant of proportionality.  The amount of tubulin getting destroyed is proportional to the pool population $n_f(t)$ and is assumed to be given by
\begin{equation}
\omega^-_{f}=\alpha^- {n_f(t)}, 
\end{equation}
where $\alpha^-$ is another constant of proportionality.

The population kinetics of the local pool is governed by the following equation 
\begin{eqnarray}
\frac{dn_f(t)}{dt}&=&\omega^+_f-\omega^-_f-\frac{dL_{fm}}{dt} 
= \alpha^+ \frac{D}{L_{fc}^2} -\alpha^- {n_f(t)}-\frac{dL_{fm}(t)}{dt}.
\end{eqnarray} 
Hence,  the steady state precursor population in the pool  is given by
\begin{equation}
n_f^{ss}=\frac{\omega^+}{\omega^-}=\frac{\alpha^+}{\alpha^-}\frac{D}{(L_{fc}^{ss})^2}~,
\end{equation}
which implies that the effective size of the local pool for flagellum $f$ is inversely proportional to the square of  $L_{fc}$.

For simplicity, we assume that the pool concentration entering into the expression of the flagellar assembly rate is the steady-state concentration of the precursors $n_{f}^{ss}$. Accordingly, 
\begin{eqnarray}
\frac{dL_{fm}(t)}{dt} &=& \frac{n_{f}^{ss}}{n_\text{max}}~J\Omega_e ~\exp \left(  -\frac{2k L_{fm}(t)}{v} \right) -(1-\rho)^2 \Gamma_r 
\label{flagella-rate-eqG}
\end{eqnarray} 

\section{Results and discussion}

\subsection{Steady-state lengths: difference between the four pairs of flagella}

The steady state membrane-bound length $L_{fm}^{ss}$ of a flagellum, whose membrane bound length $L_{fm}$ is sensed using a time of flight mechanism, is given by
\begin{equation}
L_{fm}^{ss} = \frac{v}{2k} \ell og \left(\frac{n_{f}^{ss}}{n_\text{max}} \frac{ J \Omega_e}{(1-\rho)^2 \Gamma_r }   \right) 
\label{expr-lss1}
\end{equation}.
The expression (\ref{expr-lss1}) implies that if the difference in the amount of kinesin-13 localised at the tip is insufficient to result in very significant difference in the resorption rates, i.e., $\Gamma_r$, is practically identical for all the flagella, then the local pool size $n_{f}^{ss}$ is a possible factor that can still lead to different membrane-bound lengths of different flagella. Equivalently, recasting Eq.(\ref{expr-lss1}) as 
\begin{equation}
L_{fm}^{ss}  = \frac{v}{2k} \ell og \left( \frac{1}{n_{max}}\frac{\alpha^+}{\alpha^-}\frac{D}{(L_{fc}^{ss})^2} \frac{ J \Omega_e}{(1-\rho)^2 \Gamma_r }   \right)=\frac{v}{2k} \ell og \left(\frac{ \mathcal{C}  }{(L_{fc}^{ss})^2} \frac{ J \Omega_e}{(1-\rho)^2 \Gamma_r }   \right)~~ \left[ \text{where} ~ \mathcal{C}=\frac{1}{n_{max}}\frac{\alpha^+}{\alpha^-}D \right]
\label{expr-lss2}
\end{equation}
we conclude that different values of $L_{fc}^{ss}$ of different flagella would result in the difference of their values of $L_{fm}^{ss}$. In other words, for the four pairs of flagella, the larger is the value of  $L_{fc}^{ss}$, the smaller is the value of $L_{fm}^{ss}$. This qualitative prediction of our theory is consistent with the experimentally known fact summarized in  Fig.\ref{fig_giardia}(b).  

\begin{figure*}
\includegraphics[width=0.70\textwidth]{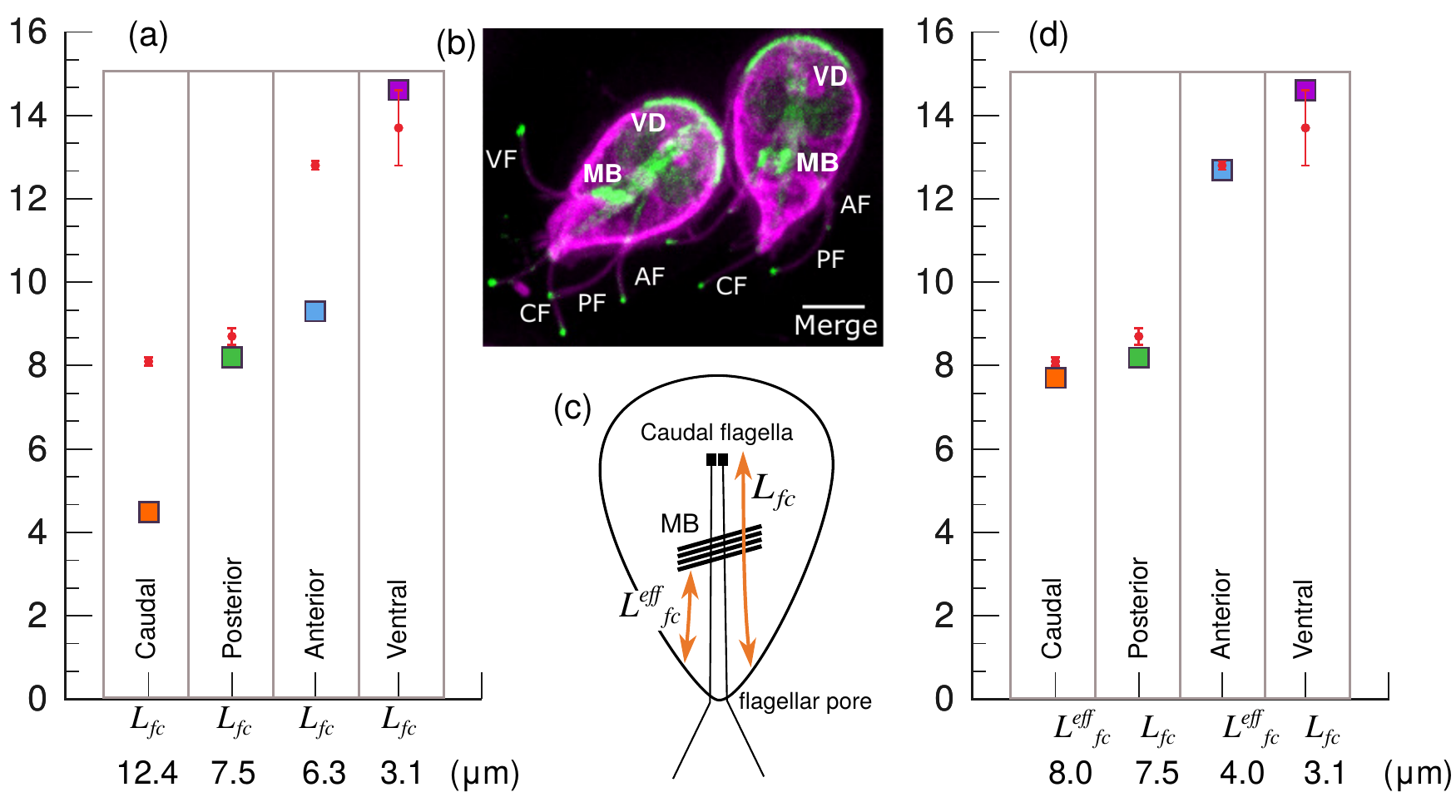}
\caption{{ {\bf Estimates of steady state length:} (a) Using the expression (\ref{expr-lss2}), we estimate the values of $L_{fm}$ for different pair of flagella using the value of $L_{fc}$ also metioned in for each flagellum type. The circular dots with the error bars are the experimentally measured values and the square dots are the estimates from the expression (\ref{fig_lss}). (b) ({Reproduced from ref.\cite{mcinally19}, which was published by McInally et al. \cite{mcinally19} under {\it Creative Commons Attribution License}}) Localisation of kinesin-13  (shown in green fluorescence) at the median body(MB) and the ventral disc (VD) suggest they are involved in depolymerizing the microtubules and releasing tubulins of MB and VD for the elongation of the axoneme. (c) As the flagellar pore of the caudal flagella are near the median body, we define a new length $L^{eff}_{fc}~(<L_{fc})$ for the caudal flagella. (d) Instead of $L_{fc}$ used in (a), we use $L^{eff}_{fc}$ for the caudal and anterior flagella in expression (\ref{expr-lss2}) to get a better estimate of $L_{fm}$.  The numerical values of the parameters used for these graphical plots are: $\rho=0.09,~ J=\rho(1-\rho)=0.819,~ v=(1-\rho)~\delta \ell=0.91\times 0.008=7.28\times 10^{-3},~ k=0.001,~ \Omega_e=0.50,~ \Gamma_r=0.0005$ and $\mathcal{C}=5.44$.}   }
\label{fig_lss}
\end{figure*}

\begin{figure*}
\includegraphics[width=0.50\textwidth]{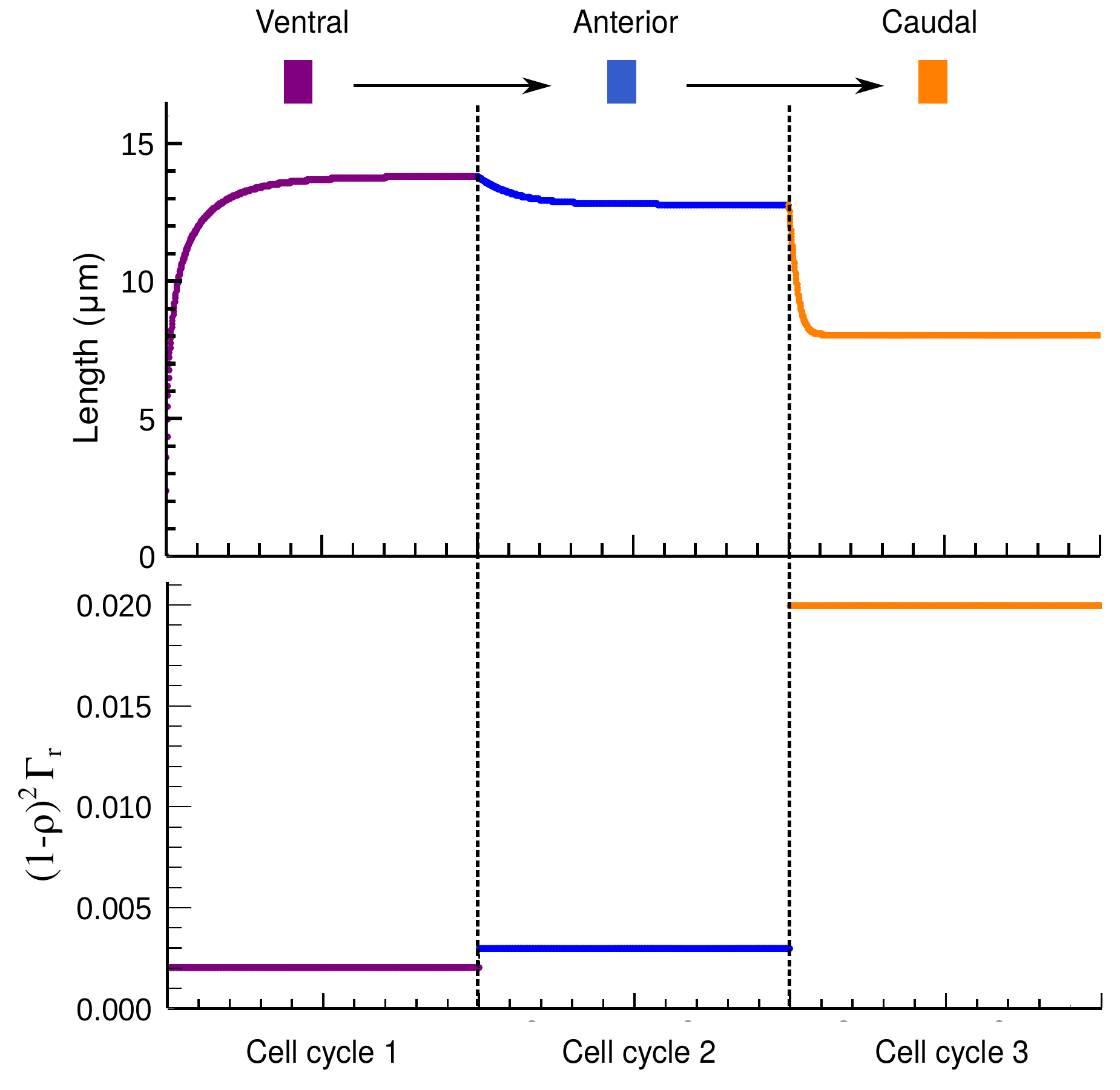}
\caption{{ {\bf Transformation of flagellum over multiple cell cycles:} (a) Equation (\ref{equation_vac}) is solved by varying the value of $\Gamma_r$  over multiple cell cycles for capturing the elongation of $L_{fm}$ of the ventral flagellum and its transformation to ventral, anterior and caudal flagellum. (b) Effective resorption rate $(1-\rho)^2 \Gamma_r(t)$ as a function of cell cycle. (Parameters: $\frac{n_{ss}}{n_{max}} J \Omega_e=0.5,~ \frac{2k}{v}=0.0032,~(1-\rho)^2 \Gamma_r(t)=0.002 ~\text{for cell cycle 1}, 0.003 ~\text{for cell cycle 2} ~\& ~ 0.02 ~\text{for cell cycle 3})$  }}
\label{fig_vac}
\end{figure*}

Next, let us make more quantitative comparison of our theoretical predictions with experimental data. 
 Using identical set of values of all the  parameters (i.e, $n_\text{max},\alpha^+,\alpha^-,D,\Omega_e,\Gamma_r,J,\rho,v,k$) for all the $f$, and the values of $L_{fc}$ for different $f$ from table in Fig.\ref{fig_giardia}(b), we estimated the value of the corresponding $L_{fm}$ from its expression (\ref{expr-lss2}). The theoretical estimates  for the posteriolateral and ventral flagella are in good agreement with the corresponding experimental values listed in the table of Fig.\ref{fig_giardia}(b) whereas our model underestimates the length of the caudal and anterior flagella (see the Fig.\ref{fig_lss}(a)). 

In order to identify the reason for the underestimation of the lengths of the caudal and anterior flagella by our model and to improve the estimates, we proceed with the following observations:  

From the table in  Fig.\ref{fig_giardia}(b) one can see that the experimentally measured $L_{fm}$ for the caudal and posteriolateral flagella are comparable whereas the length of anterior and ventral are comparable.  

(i) Localisation of kinesin-13  (shown in green fluorescence in Fig.\ref{fig_lss}(b)) at the median body(MB) and the ventral disc (VD) suggest they are involved in depolymerizing the microtobules and releasing tubulins of MB and VD for the elongation of the axoneme. 

(ii) As shown in the schematic figure of Giardia cell (Fig.\ref{fig_giardia}(a)), flagellar pore of the caudal and posteriolateral are close to the median body (MB) thereby indicating that they draw tubulins liberated from the microtubules of the MB. Similarly, the proximity of the flagellar pore of the anterior and ventral flagella to the ventral disc (VD) indicate that they seem to draw tubulins liberated from  microtubules of the VD. 

(iii) $L_{fc}^{ss}$ appears in the expression (\ref{expr-lss2}) through $n_{f}^{ss}$ in Eq. (\ref{expr-lss1}). Therefore, in principle, $L_{fc}^{ss}$ of caudal and posteriolateral flagella that draw tubulin mainly from MB should be comparable, but different from  the other two that draw tubulin mainly from the VD. A similar argument can be given in support of the comparable values of the anterior and ventral flagella that draw tubulins mainly from the VD.

Based on the observations (i)-(iii),  we propose the following:  
(a) instead of $L_{fc}^{ss}$ one should use $L^{eff}_{fc}$  (see Fig.\ref{fig_lss}(c) for the schematic description of $L_{fc}$ and $L^{eff}_{fc}$  of caudal flagella); and 
(b) the value $L_{fc}^{eff}$ for caudal flagella  comparable to $L_{fc}$ of posteriolateral flagella, whereas the value $L_{fc}^{eff}$ for the anterior flagella  comparable to $L_{fc}$ of the ventral flagella. In Fig.\ref{fig_lss}(d), the values of $L_{fc}$ and $L^{eff}_{fc}$ for each flagellum type is mentioned which are used in expression (\ref{expr-lss2}) for getting  estimates of $L_{fm}$ for different flagella. Note that the  chosen value $L_{fc}^{eff}$ for caudal flagella is not exactly equal to $L_{fc}$ of posteriolateral flagella and the value $L_{fc}^{eff}$ for the anterior flagella  is not exactly equal to $L_{fc}$ of the ventral flagella. For caudal and anterior flagella, the values of $L_{fc}^{eff}$ are varied slightly around $L_{fc}$ of posteriolateral flagella and $L_{fc}$ of the ventral flagella, respectively, in such a manner that the respective values of $L_{fm}$ predicted by the theory are a good agreement with the corresponding experimentally recorded values. Implementing this proposal, we get new estimates of the values of $L_{fm}^{ss}$ for caudal and anterior flagella which are, indeed, comparable to those observed experimentally (see Fig.\ref{fig_lss}(d)). 

It is an established fact that in monoflagellates \cite{patra21} and biflagellates \cite{patra20, marshall01}, size of the precursor pool at the flagellar base plays a role in determining the steady state length of the membrane bound flagellum. Here, we have proposed that emergence of variable pool sizes of tubulins at the bases of flagella, which occurs because of the diffusive transport of tubulins along the intracellular segments of the flagella, could be one of the driving factors behind the non-identical membrane-bound flagellar lengths  $L_{fm}^{ss}$. 

\subsection{Deposition of depolymerase at the tip leads to the transformation of flagellar length}
One important difference between the flagella of {\it Giardia} and {\it Chlamydomonas} is that in {\it Chlamydomonas} both the flagella are of same age i.e, after cell division they are assembled simultaneously. But this is not the case in {\it Giardia}. From Fig.\ref{fig_division}(a) one can see that the oldest flagella are the pair of  caudal flagella. Ventral flagella  whose cytoplasmic axonemes are assembled prior to the cell division mature during the first cell cycle by assembling the membrane bound flagellum and  growing it to the full length. Thereafter, during the successive cell cycle the pair of ventral flagella inherited by one of the daughter cell transform into a anterior flagella and finally in the next cell cycle the individual anterior flagellum inherited by each daughter cell  transform into a caudal flagellum. From Fig.\ref{fig_giardia}(b) and Fig.\ref{fig_division}(b) one can verify that initially the flagellum (ventral) attains its full length $L_{fm}$ but then the membrane bound length  gradually decrease with subsequent cell cycles (see Fig.\ref{fig_vac}(a)). As the amount of kinesin-13 is inversely proportional to $L_{fm}$, this indicates that the  deposition of kinesin-13 at the flagellar tip increase its amount with time (or cell cycle).  

For the membrane bound length, we started with the following equation 
\begin{eqnarray}
\frac{d L_{fm} (t)}{dt} &=& \frac{ n_{ss} }{n_\text{max}}J\Omega_e \exp \left(  -\frac{2k  L_{fm} (t)  }{v} \right) -(1-\rho)^2 \Gamma_r (t)
\label{equation_vac}
\end{eqnarray}
where the effect of changing amount of depolymerase at the tip is captured by a time varying $\Gamma_r(t)$. Note that we have kept the value of $n_{ss}$ constant. The value of $\Gamma_r(t)$ remains constant throughout the cell cycle but changes when the cell enters the next cell cycle.  The functional form of the effective resorption rate $(1-\rho)^2 \Gamma_r$ is depicted in Fig.\ref{fig_vac}(b).  This modification describes the  transformation of the membrane bound length of the flagellum which starts as a ventral flagellum in cell cycle 1, transforms into an anterior flagellum in cell cycle 2 and finally converts into a caudal flagellum in cell cycle 3.

Although this mechanism successfully explains the transformation  of flagellum from ventral to caudal, but it doesn't explain the nonmonotonous course of the flagellum which initially starts as posteriolateral flagellum, converts into anterior and then into caudal flagellum (see Fig.\ref{fig_division}(b)).

\section{Summary and conclusion}

How an eukaryotic cell controls the lengths of its flagella is one of the fundamental questions in biological physics 
(see ref.\cite{patrathesis,patra22} for reviews).  So far our understanding on flagellar length control and dynamics are gained by studying the biflagellate {\it Chlamydomonas} which have two flagella of equal length and age \cite{wemmer07}. Both the flagella elongate simultaneously after cell division and undergo complete resorption  prior to cell division. In contrast to this simple example, {\it Giardia} is a fascinating multiflagellate with flagella of different age and length.

In this paper we have developed a new model for flagellar length control in {\it Giardia}, an octoflagellate uni-cellular eukaryote. Two unusual features of these flagellate organism that we have studied here are (a) the relative sizes of the cytoskeletal parts and membrane-bound parts of the four pairs of flagella, and (b) the change in the identities (and concomitant lengths) of the four pairs of flagella with successive cell cycles. Our results are consistent with the experimental results available in the literature. 

In this paper we have focussed mainly on the steady-state lengths of the four pairs of flagella of {\it Giardia}. It would be interesting to explore the time-dependence of the mean lengths of these four pairs of flagella immediately after cell division till these attain their steady-state values. Several other fundamental questions in the context of flagellar length control, including the nature of the correlations in the length fluctuations of different flagella remain open for future investigations.

{\bf Acknowledgements}: This research is supported by SERB (India) through J.C. Bose National Fellowship of DC. \\

\end{document}